\documentclass[final,onefignum,onetabnum]{siamonline190516}

\usepackage{lipsum}
\usepackage{amsfonts}
\usepackage{epstopdf}
\ifpdf
  \DeclareGraphicsExtensions{.eps,.pdf,.png,.jpg}
\else
  \DeclareGraphicsExtensions{.eps}
\fi

\usepackage{datetime}
\newdateformat{monthyeardate}{%
  \monthname[\THEMONTH], \THEYEAR}

\usepackage{amsmath}
\allowdisplaybreaks
\usepackage{amssymb}
\usepackage{commath}
\usepackage{mathtools}
\usepackage{bbm}

\usepackage{color}
\usepackage{graphicx}
\usepackage[small]{caption}
\usepackage{subcaption}

\usepackage{relsize}
\usepackage{adjustbox}
\usepackage{algorithm}
\usepackage[noend]{algpseudocode}
\usepackage{booktabs}
\usepackage{tikz}

\usepackage{enumitem}
\setlist[enumerate]{leftmargin=.5in}
\setlist[itemize]{leftmargin=.5in}

\newsiamremark{remark}{Remark}
\newsiamremark{hypothesis}{Hypothesis}
\crefname{hypothesis}{Hypothesis}{Hypotheses}
\newsiamthm{claim}{Claim}

\headers{Oscillatory dynamics in the dilemma of social distancing}{Alina Glaubitz, Feng Fu}

\title{Oscillatory dynamics in the dilemma of social distancing}

\author{Alina Glaubitz\thanks{Department of Mathematics, Dartmouth College, Hanover, NH 03755, USA (\email{alina.glaubitz.gr@dartmouth.edu}, 
\email{feng.fu@dartmouth.edu})}
\and Feng Fu\footnotemark[1] \thanks{Department of Biomedical Data Science, Geisel School of Medicine at Dartmouth, Lebanon, NH 03756, USA}}

\usepackage{amsopn}

\ifpdf
\hypersetup{
}
\fi

\begin{document}

\maketitle

\begin{abstract}
    	Social distancing as one of the main non-pharmaceutical interventions can help slow down the spread of diseases, like in the COVID-19 pandemic. Effective social distancing, unless enforced as drastic lockdowns and mandatory cordon sanitaire, requires consistent strict collective adherence. However, it remains unknown what the determinants for the resultant compliance of social distancing and their impact on disease mitigation are. Here, we incorporate into the epidemiological process with an evolutionary game theory model that governs the evolution of social distancing behavior. In our model, we assume an individual acts in their best interest and their decisions are driven by adaptive social learning of the real-time risk of infection in comparison with the cost of social distancing. We find interesting oscillatory dynamics of social distancing accompanied with waves of infection. Moreover, the oscillatory dynamics are dampened with a nontrivial dependence on model parameters governing decision-makings and gradually cease when the cumulative infections exceed the herd immunity. Compared to the scenario without social distancing, we quantify the degree to which social distancing mitigates the epidemic and its dependence on individuals' responsiveness and rationality in their behavior changes. Our work offers new insights into leveraging human behavior in support of pandemic response.
\end{abstract}


\begin{subjects}
	Statistical physics, applied mathematics
\end{subjects}

\section{Introduction} 
\label{sec:introduction} 
Emerging novel zoonotic diseases, such as Zika~\cite{petersen2016zika}, Ebola~\cite{leroy2005fruit}, and the most recently COVID-19~\cite{andersen2020proximal}, have imposed great threats to global health and humanity~\cite{lloyd2017infectious}. Some of these new diseases are caused by respiratory viruses and highly contagious through proximity transmissions, and may turn into an unprecedented pandemic well before effective treatments and vaccines have been developed and widely deployed. In this case, the world may have to resort to non-pharmaceutical interventions (NPI), such as face covering and social distancing so as to mitigate disease impact before effective pharmaceutical interventions become available. However, the ultimate effectiveness of NPI measures is highly contingent on compliance and adherence, since NPI is usually not a one-off measure, but rather requires repeated, consistent adherence in order to reduce potential transmission routes of contracting the infection. From this perspective, human behavior plays an important role in impacting the course of a pandemic outbreak as well as the health outcome. 

In recent years, there has been growing interest in understanding social factors in epidemiology (see, for example, ~\cite{bauch2013social} for a brief review). In the field of behavioral epidemiology, of particular interest is the use of disease-behavior interaction models for this purpose~\cite{verelst2016behavioural}. Prior work has extensively used this framework to study how vaccine compliance can be influenced by a wide range of factors~\cite{Fu_PRSB11,Reluga_MB06,Wang_PR16}, ranging from vaccine scares~\cite{Bauch_PLoSCB12} to disease awareness~\cite{Wang_SR16}. The feedback loop between behavioral change and disease prevalence gives rise to a variety of interesting, nontrivial dynamics~\cite{salathe2008effect,funk2010modelling,fenichel2011adaptive,fu2017dueling}, e.\,g.\ the hysteresis effect~\cite{chen2019imperfect}. Among others, an important approach is combining evolutionary game theory with epidemiological models~\cite{bauch2005imitation,Fu_PRSB11,arefin2020vaccinating}. Evolutionary game theory provides a general mathematical framework for modeling behavioral changes in a population driven by both social influence and self interest. In the past decades, the approach of replicator dynamics has been commonly used to model social learning/imitation process, and particularly the spread of behavior (`social contagion'),  in a range of important real-world problems~\cite{cressman2014replicator}, from peer punishment~\cite{sigmund2001reward} over cooperation~\cite{perc2017cooperation}, ~altruistic punishment~\cite{page2000altruistic}, honesty~\cite{capraro2020honesty}, trust~\cite{kumar2020trust} and moral behavior in general~\cite{capraro2019moral} to antibiotic usage~\cite{chen2018social}. 

Unlike vaccination, social distancing effort of an individual requires repeated decisions whether or not to comply by evaluating the necessity of doing so throughout the epidemic, despite public health recommendations or even mandates~\cite{Townsend2020}. The cost of social distancing is not negligible, but rather has a huge impact on the economic status and well-being of people~\cite{nicola2020socio}. Previous work has modeled social distancing as a differential game~\cite{reluga2010game}, that is, individuals try to maximize their payoffs by adjusting their effort in social distancing (namely, the level of exposure to potential transmission routes) by comparing the risk of contracting the disease with the cost of social distancing. Their numerical results show that the collective dynamics of social distancing would approach to a steady level (i.\,e.\ a Nash equilibrium with constant effort for social distancing) without any oscillatory dynamics~\cite{reluga2010game}. While this prior study sheds useful insights for social distancing from the game theory perspective, it remains largely unknown how the rationality and the responsiveness of individuals in reacting to an epidemic would impact the compliance level of social distancing.

Social distancing is costly, yet if not optimized for timing and duration and intensity, it would lead to wasted effort~\cite{maharaj2012controlling,huberts2020optimal}. Combined with real data, the impact of social distancing can also be quantitatively assessed and optimized for past pandemics like influenza~\cite{caley2008quantifying,wallinga2010optimizing}. Noteworthy, there have been efforts to predict and quantify the effectiveness of reactive distancing on the COVID-19 pandemic, in anticipation of multiple waves of infections in the coming years~\cite{aleta2020modelling,kissler2020projecting}.

Aside from individual perspective, the optimization of disease control is often studied using optimal control theory by assuming a central social planner aiming to minimize the cost of disease outbreak~\cite{sethi1978optimal,abakuks1974optimal,abakuks1973optimal,wickwire1975optimal}. While these results are insightful from the perspective of population optima~\cite{morris2020optimal} (that is, optimized policies are complied uniformly in the population), it is challenging to attain these goals in practice due to compliance issues.

To shed light on driving factors of compliance levels of social distancing, here we take into account important aspects of human decision-making -- bounded rationality~\cite{simon1990bounded} and loss aversion~\cite{tversky1992advances} -- which is informed by the real-time disease prevalence, and prompted by peers' choice. We incorporate into the epidemiological process with an evolutionary game dynamics of social distancing behavior. Individuals decide on whether or not to commit to social distancing by weighing the risk of infection with the cost of social distancing. The responsiveness parameter in our model modulates the relative time scale of individuals revisiting their social distancing decisions, as compared to the pace of an unfolding epidemic. We introduce bounded rationality that individuals are not necessarily using the best response but rather with some probability of changing their behavior. 

In this work, we find an interesting \emph{oscillatory tragedy of the commons} in the collective dynamics of social distancing. Individuals are inclined to social distancing when the disease prevalence is above a threshold that depends on the transmissibility of the disease and the relative cost of social distancing versus contracting the disease. As the epidemic curve is being flattened, individuals consequently feel more safe not to practice social distancing, thereby causing the decline in the compliance of social distancing and further resulting in a resurgence of disease outbreaks in the population. Even though such reactive social distancing is hardly able to help reach the optimality of disease mitigation, it can avoid the overshooting of infected individuals which typically happens in an susceptible-infected-recovered (SIR) model in the absence of any interventions. We also find nontrivial dependence of the effectiveness of social distancing, measured by the fraction of susceptible individuals who would become infected without social distancing, on model parameters governing individuals' rationality and responsiveness. 

\section{Model and Methods}

\subsection{Model}
Our model is basically a combination of the classical SIR model with the replicator equation: In a well-mixed infinite population each individual is either susceptible, infected or recovered. 
Moreover, each susceptible individual can at each time choose to either practice social distancing or not to practice social distancing. If the individual practices social distancing they cannot become infected. If they do not practice social distancing they become infected in an encounter with an infected with probability $\beta>0$. At each time an infected recovers with probability $\gamma>0$. 

\begin{figure}[h!]
    \centering
    \includegraphics[width=2.5in]{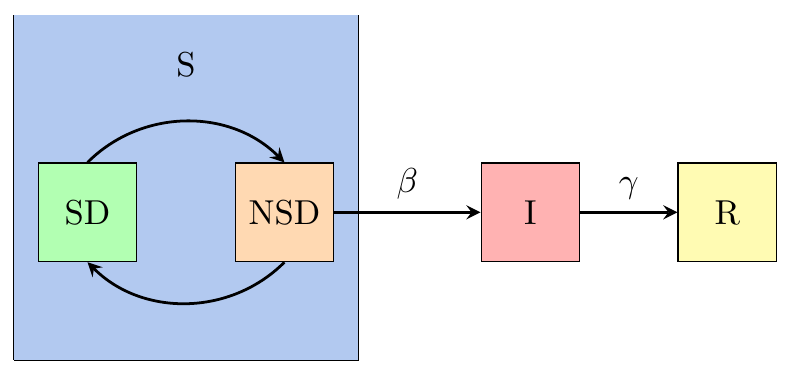}
\caption{Schematic of our model. In contrast to the SIR model, we divide the susceptible population in those that practice social distancing (and cannot be infected) and those that do not practice social distancing (and can be infected). The dynamics of the amount of people practicing social distancing is given by a replicator equation. Here as well as in later figures, we refer to those practicing social distancing as SD and those that do not as NSD.}
\label{fig:schematic}
\end{figure}

\begin{figure}[h!]
    \centering
    \includegraphics[width=4in]{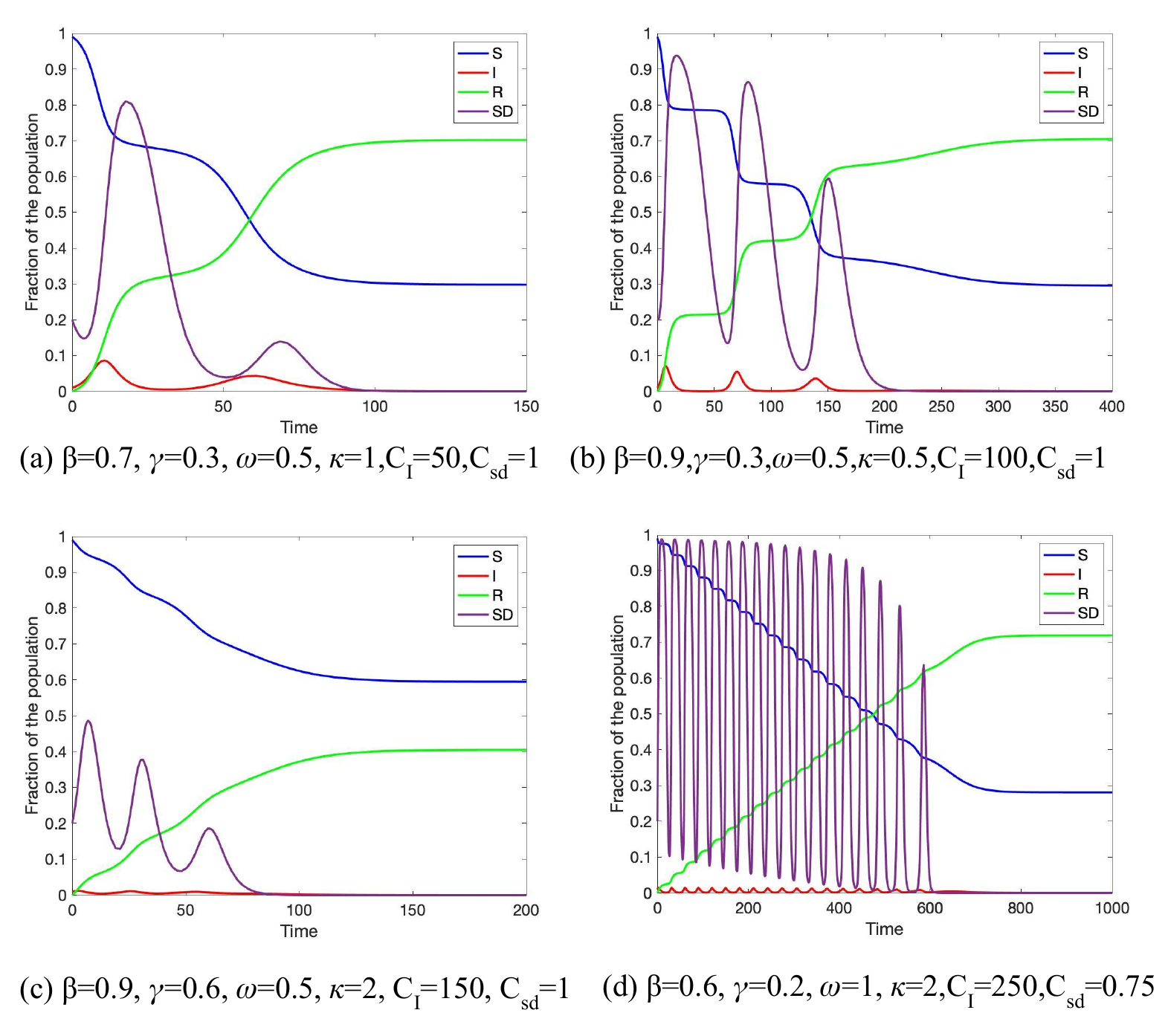}
\caption{Model \eqref{eq:SIR_SD} for different parameters with $I_0=0.01,S_0=0.99,\mathcal{E}_0=0.2$. SD Here, as well as for all other figures, we used the MATLAB method ode23, which is an implementation of the Bogacki-Shampine method---an explicit Runge-Kutta (2,3) pair.}
\label{fig:i_star}
\end{figure}

We denote the proportion of susceptible individuals at time $t$ by $S(t)$, the proportion of infected by $I(t)$, and the proportion of removed by $R(t)$. Furthermore, we denote by $\mathcal{E}(t)$ the proportion of susceptible individuals that practice social distancing. We denote the initial conditions by $I_0 = I(0), S_0=S(0)$ as well as $\mathcal{E}_0 = \mathcal{E}(0)$.
A susceptible individual determines his strategy based on a cost-benefit analysis. Hence,
by $\pi_{\text{sd}}$ we denote the payoff of social distancing, and by $\pi_{\text{nsd}}$ the payoff of no social distancing. In our model the perceived cost of social distancing is $C_{\text{sd}}>0$ at each time $t$. Thus, we have 
\[
    \pi_{\text{sd}} = - C_{\text{sd}}.
\]
$\pi_{\text{nsd}}$ depends on two factors: the perceived cost of infection that we denote by $C_{\text{I}}>0$ and the risk of infection. The risk of infection in time $(t,t+1)$ without social distancing is given by 
\[
    1 - \exp\left(-\beta \int_t^{t+1} I(\tau) \mathrm{d}\,\tau\right) \approx 1 - \exp\left(-\beta I(t)\right).
\]
Therefore, the payoff of not socially distancing is given by
\[
    \pi_{\text{nsd}} = -C_{\text{I}} (1-\exp\left(-\beta I(t)\right)).
\]
Hence, the dynamics of our model are given by the following system of ordinary differential equations (ODEs):
\begin{align}\label{eq:SIR_SD}
\begin{split}
    \dot{S}(t) &= -\beta (1 - \mathcal{E}(t)) S(t) I(t)\\
    \dot{I}(t) &= \beta (1 - \mathcal{E}(t)) S(t) I(t) - \gamma I(t) \\
    \dot{R}(t) &= \gamma I(t)\\
    \dot{\mathcal{E}}(t) &= \omega \mathcal{E}(t) (1-\mathcal{E}(t)) \tanh\left(\frac{\kappa}{2}\left(-C_{\text{sd}} + C_{\text{
    I}} \left(1 - e^{-\beta I(t)}\right)\right)\right)
\end{split}
\end{align}
Here, $\omega$ is a responsiveness parameter, determining the time scale for updating the social distancing behavior. $\kappa$ is a rationality parameter. For large $\kappa$ individuals change their strategy if the payoff of the other strategy is larger. For small $\kappa$ only a fraction of the susceptible individuals depending on the difference in payoff change their strategy. The behavior of this model is illustrated in Figure \ref{fig:i_star} for different parameters. For this figure as well as for all other figures, we used the MATLAB method ode23, which is an implementation of the Bogacki-Shampine method---an explicit Runge-Kutta (2,3) pair.

\subsection{Perfect Adaption}

In Model \eqref{eq:SIR_SD} the dynamics of social distancing $\mathcal{E}$ change to direct the amount of infected $I$ towards the amount where $\pi_{\text{sd}} = \pi_{\text{nsd}}$, i.\,e.\ towards $I^*$. Assuming that this adaption works perfectly, we obtain the following model.

\begin{figure}[h!]
    \centering
    \includegraphics[width=2.5in]{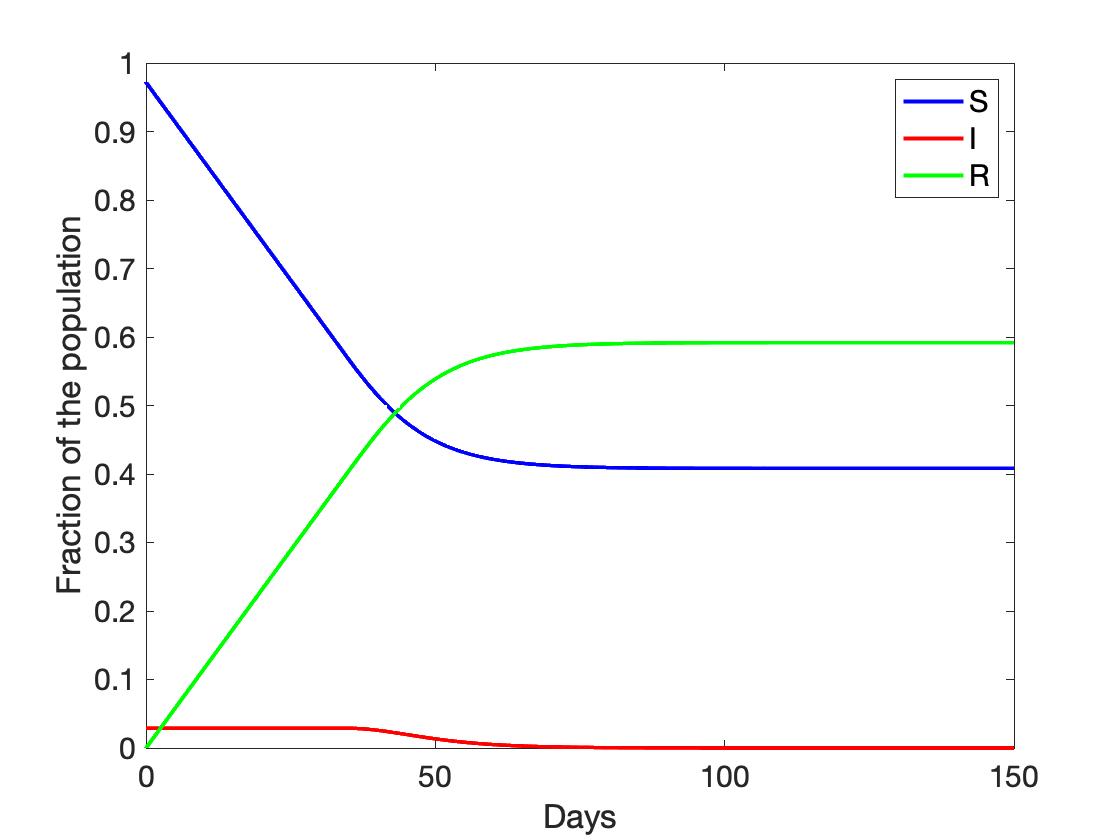}
    \caption{Perfect Adaption with $\beta = 0.7, \gamma = 0.4, C_d = 0.75, C_I = 50$. Here, $I(t) = I^*$ until herd immunity occurs, i.\,e.\ $S < \frac{\gamma}{\beta}$. From that point on, the model is a classical SIR model with $I_0 = I^*,\ R_0 = \gamma I^* t^* = 1 - \frac{\gamma}{\beta} - I^*$.}
    \label{fig_perfect_adaption}
\end{figure}

This model is given by the ODEs
\begin{align}\label{eq:perfect_adaption}
\begin{split}
    \dot{I}_{\text{PA}}(t) &= \begin{cases}
            0,&  t < t^* \\
            \beta I(t) S(t) - \gamma I(t), & t>t^*
        \end{cases} \\
    \dot{R}_{\text{PA}}(t) &= \gamma I(t)
\end{split}
\end{align}
with $S(t) = 1 - I_{\text{PA}}(t) - R_{\text{PA}}(t)$, $t^* = - \frac{\gamma - \beta + \beta I^*}{\beta I^* \gamma}$, and with initial condition 
\[
    I_{\text{PA}}(0) = I^*, \qquad R_{\text{PA}}(0) = 0.
\]
Then, the total amount of people that get infected $R_{\text{PA}}(\infty)$ is given by
\[
    R_{\text{PA}}(\infty) = \frac{\gamma}{\beta} W\left( - \exp\left( -\frac{I^* \beta}{\gamma} - 1 \right) \right) + 1,
\]
where $W$ denotes the Lambert $W$ function.
Thus, in the case of perfect adaption, we can achieve
\[
    R_{\text{PA}}(\infty) \to 1 - \frac{\gamma}{\beta}
\]
by choosing $\frac{C_d}{C_I} \to 0$. We want to use this model of perfect adaption to understand how the total amount of infected $R(\infty)$ in Model \eqref{eq:SIR_SD} depends on the parameters $C_{\text{sd}}, C_{\text{I}}, \omega,\kappa$.

\section{Results}

\subsection{Oscillatory Tragedy of the Commons}

\begin{figure}[h!]
    \centering
    \includegraphics[width=4in]{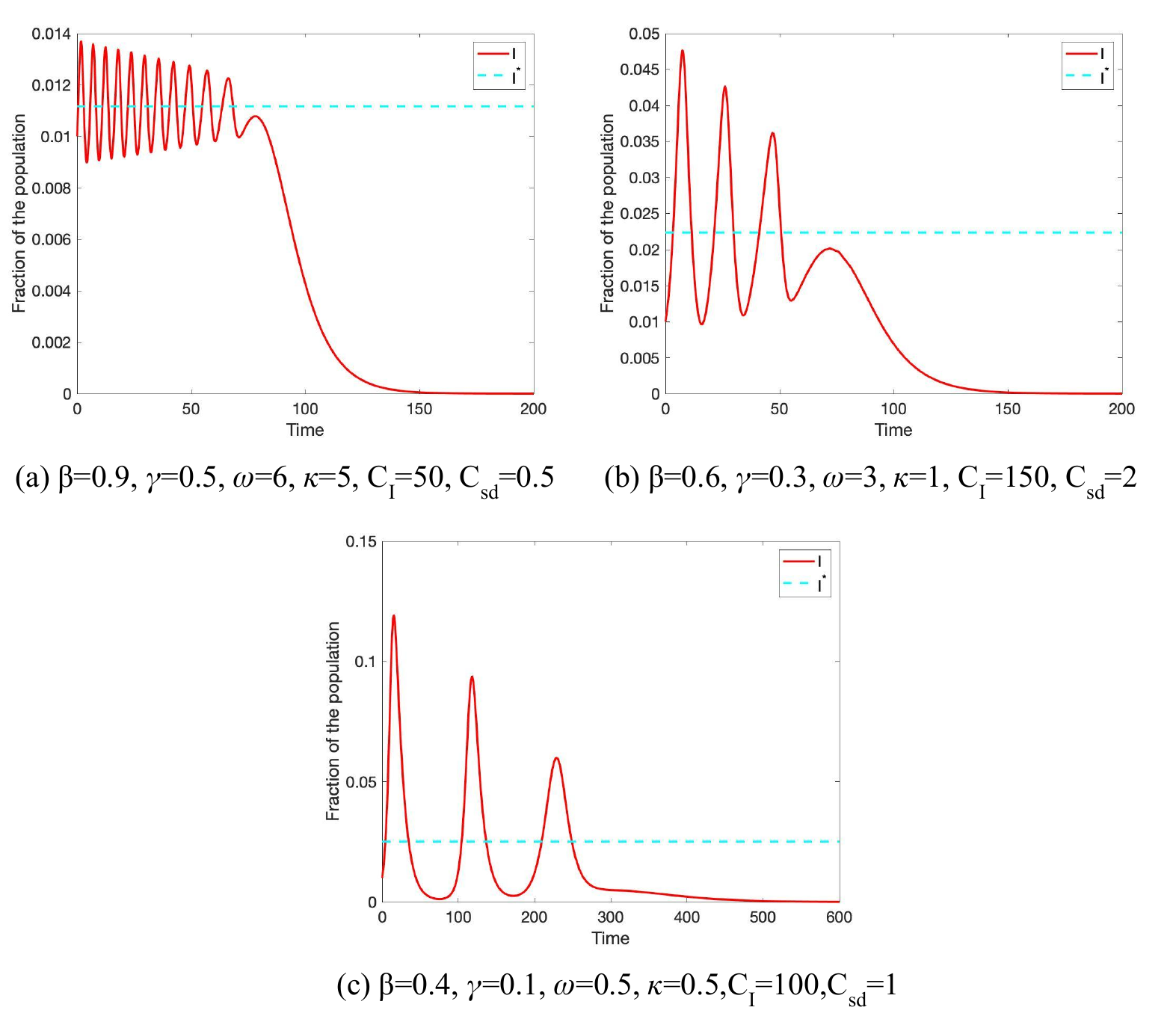}
    \caption{Oscillations of the number of infected around $I^*$ in Model \eqref{eq:SIR_SD} with initial condition $I_0 = 0.01, S_0 = 0.99, \mathcal{E}_0 = 0.3$. In each instance, $I$ oscillates around $I^*$ with decreasing amplitude until the peak is smaller than $I^*$. We see that the number and amplitude of oscillations depends on the parameters $\omega,\kappa,C_{\text{sd}},C_{\text{I}}$ as well as $\beta$ and $\gamma$.}
    \label{fig:IStar}
\end{figure}

In Model \eqref{eq:SIR_SD}, the cost of social distancing and no social distancing are equal at time $t$ if
\[
    I(t) = I^* := - \frac{1}{\beta} \log\left( 1 - \frac{C_{\text{sd}}}{C_{\text{I}}} \right).
\]
If $I>I^*$, then $\mathcal{E}$ is increasing. If $I<I^*$ then $\mathcal{E}$ is decreasing. On the other hand, if $\mathcal{E}$ is sufficiently large, this causes a decrease in $I$ and if $\mathcal{E}$ is sufficiently small this causes an increase in $I$. If the amount of infections is high, people are more aware of the disease and practice social distancing. As soon as the amount of infections is small again, this awareness fades and people do not feel the need to practice social distancing anymore. As a result, more people become infected again leading to a higher awareness and more people practicing social distancing. We refer to this feedback loop as \textit{oscillatory tragedy of the commons}.
Instead of high compliance to social distancing until the disease has died of, we find a decrease in individuals practicing social distancing when the amount of infected is sufficiently small. This then causes another rise of infections. We can observe this in Model \eqref{eq:SIR_SD} as $I$ oscillates around $I^*$ with decreasing amplitude until the peak of the oscillations is smaller than $I^*$ (see Figure \ref{fig:IStar}).

\subsection{Social Distancing Saves Lives} \label{lifes}

\begin{figure}[h!]
    \centering
    \includegraphics[width=4in]{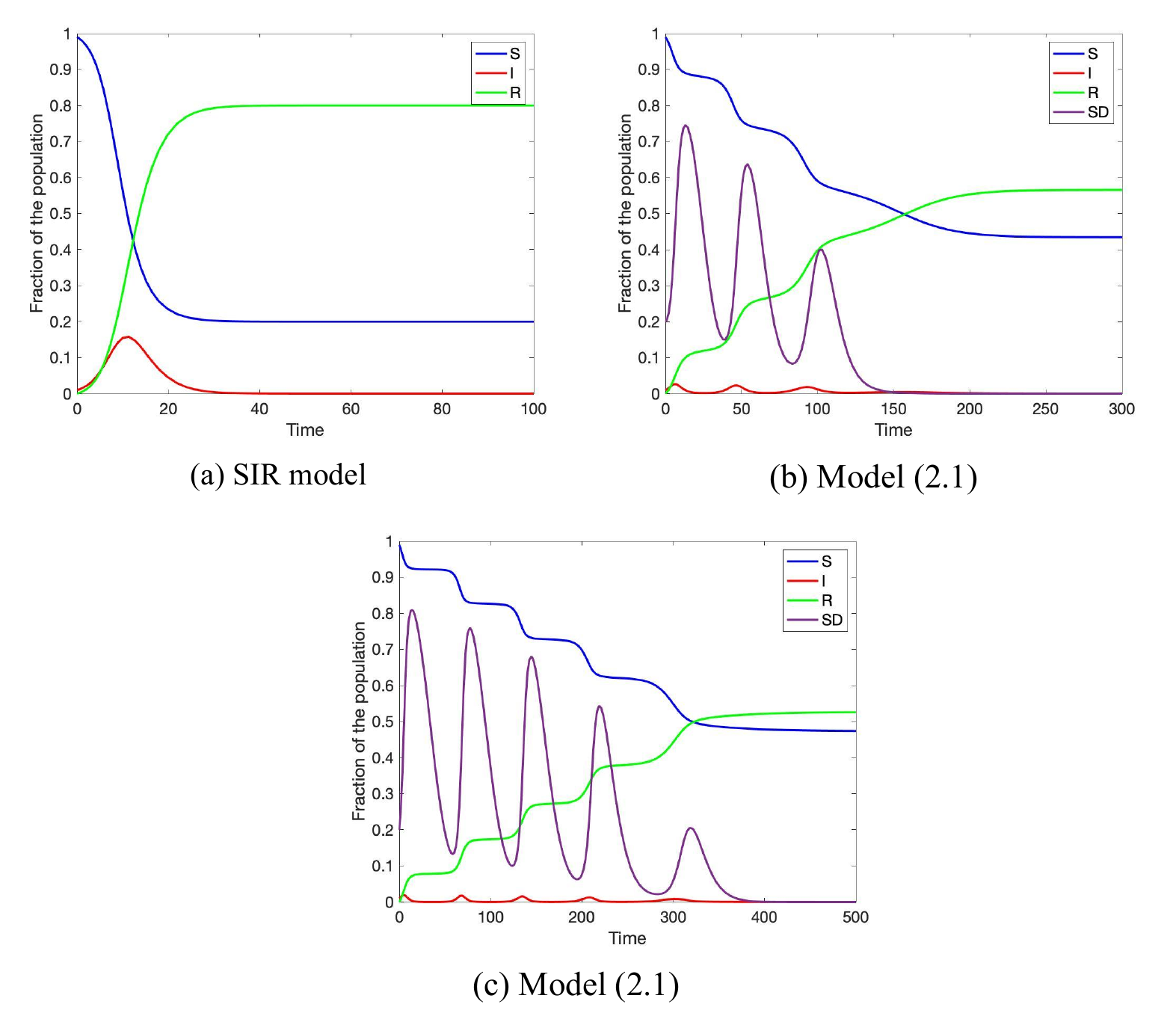}
    \caption{Comparison of the SIR model with Model \eqref{eq:SIR_SD} for $\beta = 0.8, \gamma = 0.4$ with $I_0 = 0.01, S_0 = 0.99, \mathcal{E}_0=0.2 ,C_{\text{sd}} =0.75$ and (b) $C_{\text{I}} =100, \omega = 1, \kappa =2$ respectively (c) $C_{\text{I}} =250, \omega = 0.5, \kappa =0.5$. We clearly see here that social distancing reduces the total amount of infections.}
    \label{fig:SIR_vs_SD}
\end{figure}

When comparing Model \eqref{eq:SIR_SD} to the SIR model, we immediately note that the total number of infections can be significantly smaller with social distancing (see Figure \ref{fig:SIR_vs_SD}). Essentially, this means that voluntary social distancing can significantly reduce the total amount of infections $R(\infty)$. However, we also note that infections after the first wave of infection only emerge due to the \textit{oscillatory tragedy of the commons}. If social distancing was practiced until $I=0$, we would have a much smaller $R(\infty)$.

\begin{figure}[h!]
    \centering
    \includegraphics[width=4in]{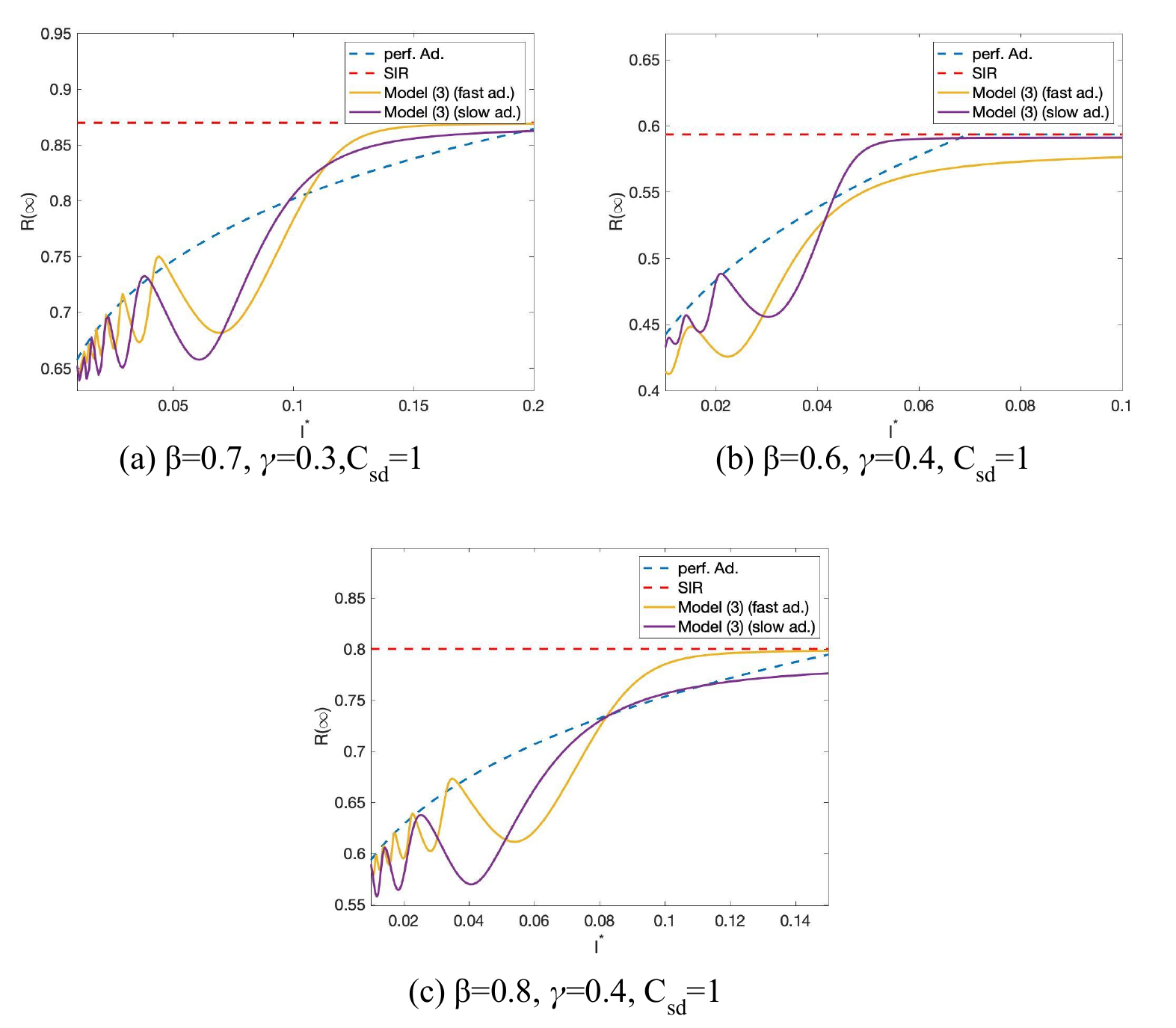}
    \caption{Comparison of the total number of infected for the SIR model, perfect adaption and Model \eqref{eq:SIR_SD} with initial values $I_0 = 0.01, S_0 = 0.99, \mathcal{E}_0=0.2$, and (a) $\omega = 1.5$ and  $\kappa = 0.5$ respectively $\omega = 2$ and $\kappa = 1$, (b) $\omega = 1$ and $\kappa = 0.5$ respectively $\omega = 1$ and $\kappa = 2$, (c) $\omega = 1$ and $\kappa = 0.5$ respectively $\omega = 1$ and $\kappa = 2$. Clearly, in perfect adaption and Model \eqref{eq:SIR_SD} the total amount of infections can be significantly smaller than in the SIR model. We see that also in Model \eqref{eq:SIR_SD} the amount of infections tends to grow if $I^*$ grows. This will be discussed in Subsection \ref{costs}. Moreover, small adaption parameters $\omega$ and $\kappa$ in Model \eqref{eq:SIR_SD} can significantly reduce infections compared to perfect adaption as well as larger adaption parameters. This will be discussed in Subsection \ref{omega}.}
    \label{fig:SD_saves_lives}
\end{figure}

Perfect adaption and Model \eqref{eq:SIR_SD} significantly reduce the total amount of infections compared to the SIR model. This is especially apparent for small $I^*$ and slow adaption, i.\,e.\ small $\omega$ and $\kappa$. This is illustrated in Figure \ref{fig:SD_saves_lives}.

Essentially, the explanation for this behavior relates to herd immunity.
Social distancing flattens the curve. Instead of one large wave of infections as in the SIR model, in Model \eqref{eq:SIR_SD} we can have several waves of infection with smaller peaks. An example of this is illustrated in Figure \ref{fig:SIR_vs_SD}. In the SIR model, herd immunity occurs if 
\[
    S < \frac{\gamma}{\beta}.
\]
Thus, $I$ is increasing until $S = \frac{\gamma}{\beta}$ and then is monotonically decreasing. Even though, we have achieved some kind of herd immunity at this point, the high number of infected $I$ still causes a high amount of new infections after herd immunity. Thus, the total amount of infections $R(\infty)$ is significantly larger than needed to obtain herd immunity.
In Model \eqref{eq:SIR_SD}, the dynamics are much more complicated. However, what remains as in the SIR model, is that as soon as (or at latest at this point) $S < \frac{\gamma}{\beta}$ the amount of infected $I$ is monotonically decreasing, since then we have 
\[
    \dot{I} = \beta (1 - \mathcal{E}) SI - \gamma I \leq \beta SI - \gamma I < 0.
\]
We denote the amount of infected when herd immunity is obtained by $I_{\text{HI}}$. With social distancing $I_{\text{HI}}$ can become significantly smaller since social distancing significantly reduces the amount of infection. Other factors influencing the amount of new infection after herd immunity are the amount of recovered when herd immunity is achieved (denoted by $R_{\text{HI}}$) as well as the amount of people practicing social distancing. $R_{\text{HI}}$ can be much larger when social distancing is practiced due to the spread of infections over a longer time period. People that practice social distancing further reduce the amount of new infections.

\begin{figure}[h!]
    \centering
    \includegraphics[width=4in]{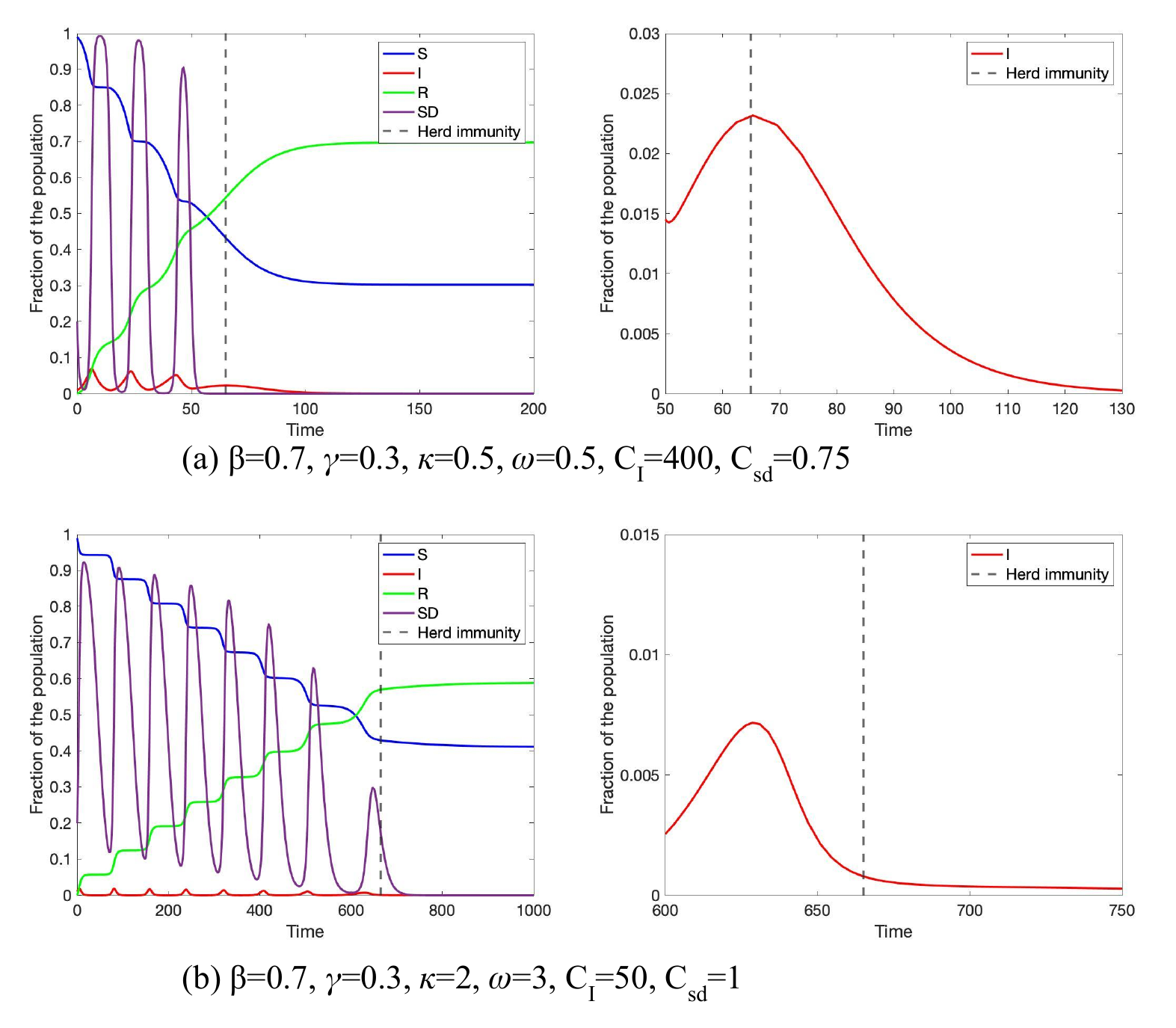}
    \caption{Occurrence of herd immunity for different parameters in Model \eqref{eq:SIR_SD}. In both examples, herd immunity occurs for $S < \frac{\beta}{\gamma}$. However, for the parameters in (a), we have significantly less infected at this point than in (b). In (a) the small amount $I_{\text{HI}}$ causes much less new infections than the larger $I_{\text{HI}}$ in (b). Therefore, the total amount of infections for (a) is much smaller than for (b).}
    \label{fig:herdImmunity}
\end{figure}

Together, all these factors cause a significant decrease in new infections after herd immunity is achieved. Since small $I_{\text{HI}}$ mostly coincides with large $R_{\text{HI}}$ as well as high $\mathcal{E}$, we focus on $I_{\text{HI}}$ here.
When choosing the parameters $\omega, \kappa, C_{\text{sd}}, C_{\text{I}}$ such that $I_{\text{HI}}$ is small, $R_{\text{HI}}$ and many people practice social distancing, we can even achieve $R(\infty)$ to be near the herd immunity threshold $1- \frac{\gamma}{\beta}$. For an illustration of this, see Figure \ref{fig:herdImmunity}.

This also explains why perfect adaption causes larger total amounts of infections than Model \eqref{eq:SIR_SD}. For perfect adaption, herd immunity is obtained for $t = t^*$ with $I = I^*$ while $I_{\text{HI}}$ often is significantly smaller in Model \eqref{eq:SIR_SD}. In particular, if we have $I > I^*$ while $S$ is close to $\frac{\gamma}{\beta}$, small increases in $\mathcal{E}$ cause a decrease in $I$. 
Hence, $I_{\text{HI}} < I^*$ in most cases.
For instance, in Figure \ref{fig:herdImmunity}a we have $I^* \approx 0.00268$ while $I_{\text{HI}}$ can be much smaller in Model \eqref{eq:SIR_SD}. However, we once again remember the \textit{oscillatory tragedy of the commons}, i.\,e.\ that higher compliance to social distancing when $I$ is small could lead to much smaller $R(\infty)$. 

Next, we want to analyze how the perceived cost of social distancing $C_{\text{sd}}$ as well as the perceived cost of infection $C_{\text{I}}$ influence the total amount of infections $R(\infty)$.


\subsection{Larger Cost of Infection and Smaller Cost of Social Distancing Reduce Infections} \label{costs}

\begin{figure}[h!]
    \centering
    \includegraphics[width=4in]{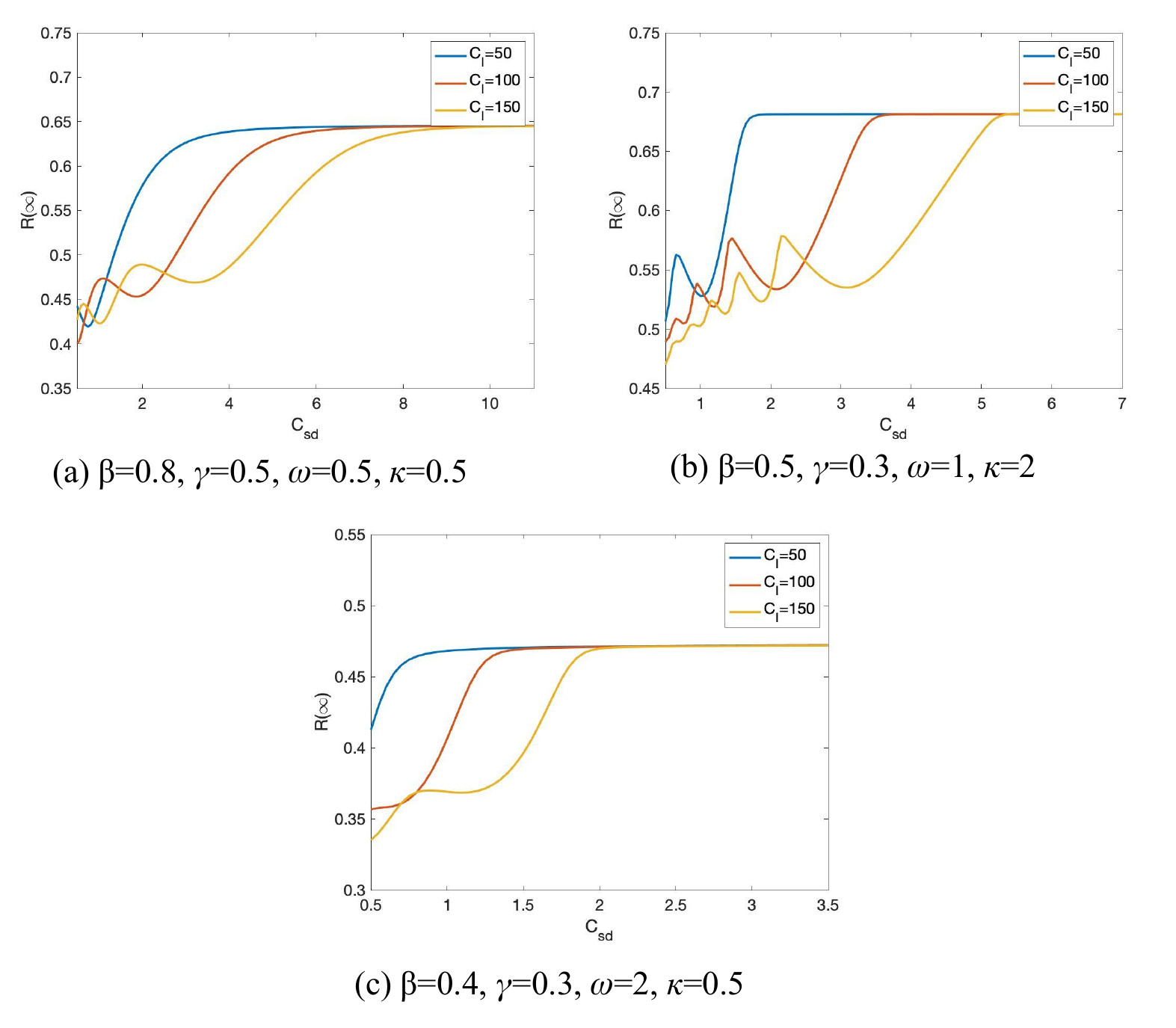}
    \caption{Total number of infections dependent of the cost of social distancing. We compare Model \eqref{eq:SIR_SD} for different costs of social distancing. Here, we have $I_0 = 0.01, S_0 =0.99, \mathcal{E}_0 = 0.3$. In all three scenarios, we see that despite small oscillations in $R(\infty)$ the number of infections is overall increasing in $C_{\text{sd}}$. Moreover, we see that for large $C_{\text{I}}$ the total number of infections can be significantly smaller, i.\,e.\ the tendency that higher costs of infection decrease the number of infections. Moreover, larger $C_{\text{I}}$ as well as larger $\kappa$ and $\omega$ cause a higher number of smaller oscillations.}
    \label{fig:csd}
\end{figure}

As one might expect, if the cost of infection $C_{\text{I}}$ increases or the cost of social distancing $C_{\text{sd}}$ decreases, this induces an increase in the amount of people practicing social distancing and thereby a decrease in infections. This behavior becomes quite apparent in Figure \ref{fig:csd}. 

\begin{figure}[h!]
    \centering
    \includegraphics[width=4in]{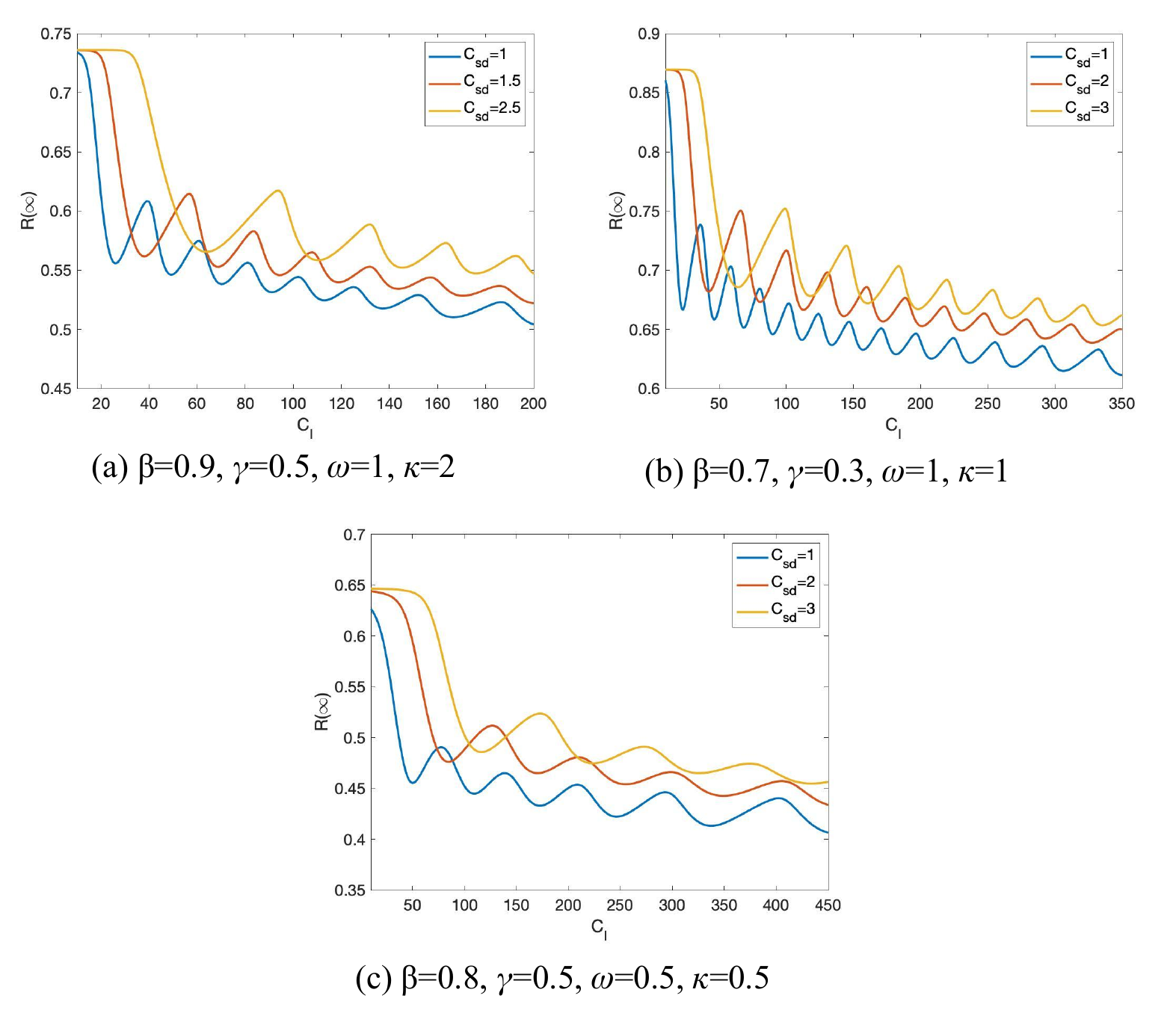}
    \caption{Total number of infections dependent of the cost of infection. We compare Model \eqref{eq:SIR_SD} for different costs of social distancing (with inital values $I_0 = 0.01, S_0 =0.99, \mathcal{E}_0 = 0.2$). $R(\infty)$ shows oscillations with decreasing amplitude. These oscillations appear to be leveling of approximately at $R_{\text{PA}}(\infty)$. We see the tendency that $R(\infty)$ decreases for increasing $C_I$. Moreover, $R(\infty)$ is smaller here again for smaller $C_{\text{sd}}$.}
    \label{fig:CI}
\end{figure}

We can observe a similar tendency for $C_{\text{I}}$. Though, here we have larger oscillations in the total size of infections. These oscillations decrease in their amplitude and level off at $R_{\text{PA}}(\infty)$. An example for this behavior can be seen in Figure \ref{fig:CI}.

\begin{figure}[h!]
    \centering
    \includegraphics[width=5in]{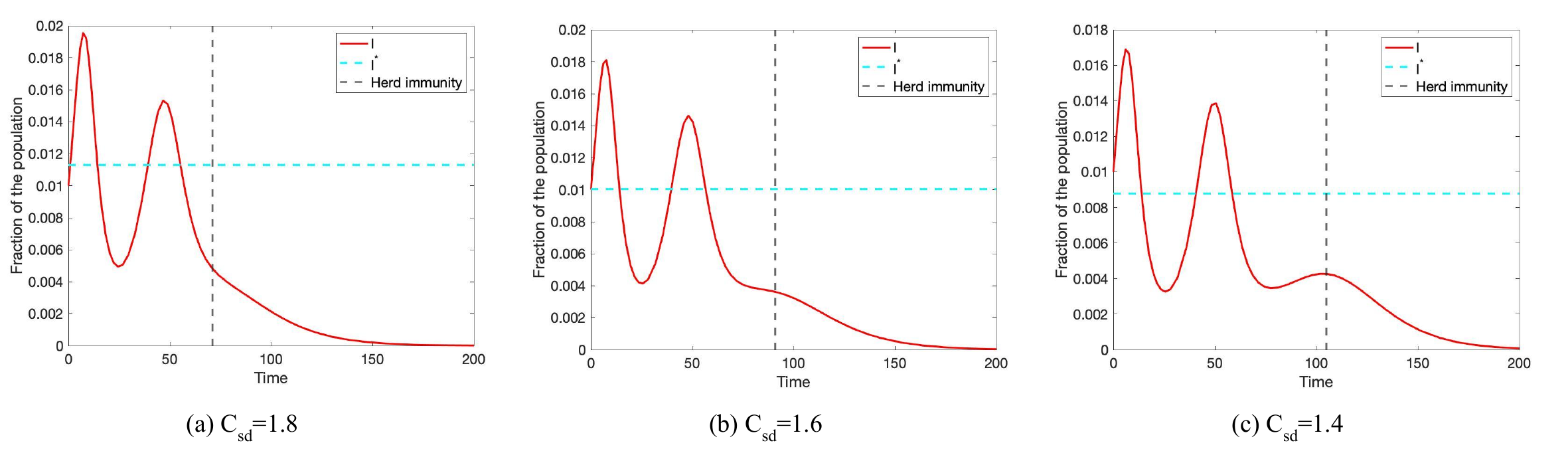}
    \caption{The amount of infected when herd immunity is obtained in Model \eqref{eq:SIR_SD} with $\beta = 0.8, \gamma = 0.5, \omega = 0.5, \kappa = 0.5,C_{\text{I}} = 200$, initial condition $I_0 = 0.01, S_0 = 0.99, \mathcal{E}_0 = 0.15$ and different $C_{\text{sd}}$.  For (a) $C_{\text{sd}} = 1.8$, we have two waves of infection and herd immunity is obtained after approximately 70 days. We have $I_{\text{HI}} \approx 0.0045$ and overall we have a proportion of $R(\infty) \approx 0.4418$ infections. For (b) $C_{\text{sd}} = 1.6$, the waves of infection decrease in size. Thus, herd immunity is obtained after a longer period of time ($t=90$). At the time, we get herd immunity, we have a smaller amount of infections ($I_{\text{HI}} \approx 0.0035$) and overall we have $R(\infty) \approx 0.4389$ infections. However, we see the beginning of the development of a third wave  of infection. For (c) $C_{\text{sd}} = 1.4$, we see the effect of this third wave of infection. Herd immunity here is obtained after 105 days with an amount of infected $I_{\text{HI}} \approx 0.0042$. Even though, we have a smaller $I^*$ here, the third wave of infection causes an increase in overall infections. Here, we have $R(\infty) \approx 0.4442$ and thus even more infections than for $C_{\text{sd}} = 1.8$.}
    \label{fig:expl_waves}
\end{figure}

An explanation for this behavior is connected to the observation made in Subsection \ref{lifes} that is illustrated in Figure \ref{fig:herdImmunity}. To reduce $R(\infty)$, (among other factors) $I_{\text{HI}}$ has to be small. One way to achieve this is to reduce $I^*$, the threshold that $I$ oscillates around. Therefore, smaller $C_{\text{sd}}$ as well as larger $C_{\text{I}}$ tend to cause a decrease in $R(\infty)$. However, this does not yet explain how the oscillations occur. For this purpose, we have a look at Figure \ref{fig:expl_waves}. Here, we see that reducing $I^*$ has two opposing effects on $R(\infty)$:
\begin{itemize}
    \item[(i)] A decrease in $I^*$ causes a decrease in the size of the waves of infections and thus a decrease in $I_{\text{HI}}$. This causes a decrease in $R(\infty)$.
    \item[(ii)] When decreasing $I^*$ too much, this can lead to the development of a new wave of infections. When this occurs, we have an increase in $I$, before herd immunity is obtained. This causes a larger $I_{\text{HI}}$. Therefore, we have an increase in the total amount of infections $R(\infty)$ when a new wave of infections develops. 
\end{itemize}

This leads to the oscillations, that we observed in Figure \ref{fig:CI}. When reducing $C_{\text{sd}}$ we first see a decrease in $R(\infty)$ (caused by smaller waves of infection and a smaller $I_{\text{HI}}$) followed by an increase (induced by a new wave of infections that leads to an increase in $I_{\text{HI}}$).

\subsection{Faster Responses and Higher Rationality Increase Infections} \label{omega}

\begin{figure}[h!]
    \centering
    \includegraphics[width=4in]{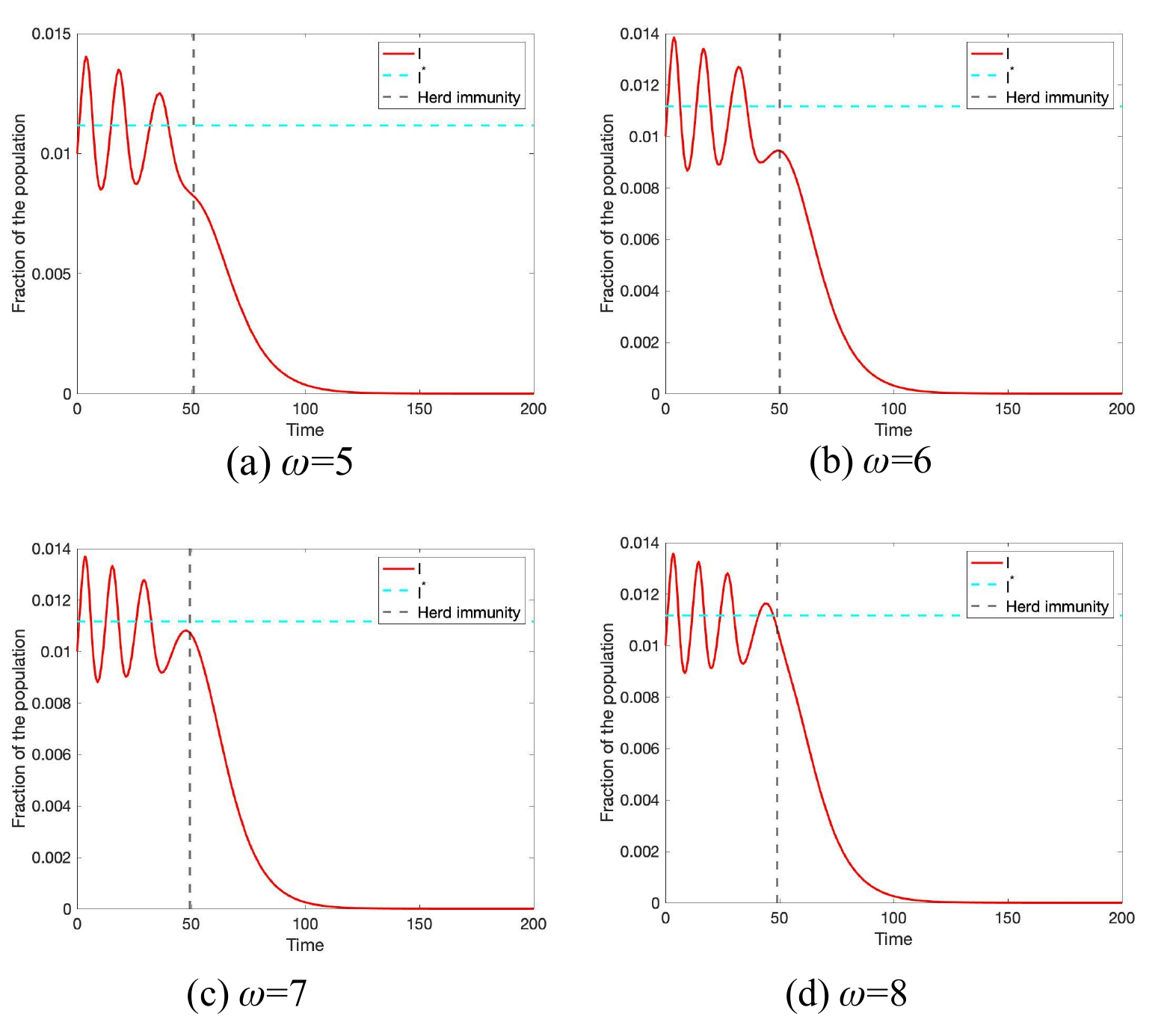}
    \caption{Development of an oscillation in the total number of infections. In Model \eqref{eq:SIR_SD} with $\beta=0.9,\gamma = 0.6, \kappa = 0.5,  C_{\text{I}}=100,C_{\text{sd}}=1$ and initial condition $ I_0 = 0.01, \mathcal{E}_0 = 0.2$, we observe the amount of infected $I$. As can be seen in Figure \ref{fig:omega}, $R(\infty)$ has a local minimum in this case approximately at $\omega = 5$. In (a) we have $\omega = 5$, herd immunity is achieved for $t \approx 50$ and at this time, we have $I_{\text{HI}} \approx 0.0083$. The total amount of infections is given by $R(\infty) \approx 0.4316$. For (b) $\omega = 6$ herd immunity is achieved at approximately the same time, with $II_{\text{HI}} \approx 0.0094$ and total amount of infections $R(\infty) \approx 0.4385$. For (c) $\omega = 7$ herd immunity is achieved at approximately the same time, with $I_{\text{HI}} \approx 0.0106$ and total amount of infections $R(\infty) \approx 0.4432$. Finally, for (c) $\omega=8$ herd immunity is achieved at approximately the same time, with $I_{\text{HI}} \approx 0.0101$ and total amount of infections $R(\infty) \approx 0.4395$.
    For increasing $\omega$, we observe that the first three waves of infection decrease in duration and intensity. However, a fourth wave of infection develops. This leads to an increase in the amount of infections when herd immunity is achieved and therefore an increase in the total amount of infections for $\omega = 6$ respectively $\omega =7$ compared to $\omega =5$. For $\omega = 8$, however, herd immunity is only obtained after the fourth wave of infection is already decreasing again. Therefore, we have a decrease in the total amount of infections for $\omega = 8$ compared to $\omega = 7$. Note also that the amount of infected when herd immunity is achieved, is in all cases smaller than $I^*$. If we have $I > I^*$ while $S$ is close to $\frac{\gamma}{\beta}$, small increases in $\mathcal{E}$ cause a decrease in $I$. Hence, $I$ is mostly smaller than $I^*$ when herd immunity occurs. This also induces that $R(\infty)$ is mostly smaller than $R_{\text{PA}}(\infty)$ as can also be seen in Figure \ref{fig:omega}. }
    \label{fig:waves_omega}
\end{figure}

Two other important factors determining $R(\infty)$ are the responsiveness $\omega$ and the rationality parameter $\kappa$. In Model \eqref{eq:SIR_SD}, a larger $\omega$ causes faster adaption of social distancing to the amount of infected. This has two opposing effects. 
\begin{itemize}
    \item[(i)] On the one hand, faster adaptions causes a decrease in the duration of the waves of infection with smaller maxima and larger minima.
    \item[(ii)]  On the other hand, these smaller waves of infection can cause the development of another wave of infection. In particular, if herd immunity is obtained before this new wave recedes, this leads to an increase in $I_{\text{HI}}$. Thus, leading to an increase of infections.
\end{itemize}
 Overall, we thus have oscillations in $R(\infty)$ depending on $\omega$.  An example of this is illustrated in Figure \ref{fig:waves_omega}. 
 
\begin{figure}[h!]
    \centering
    \includegraphics[width=4in]{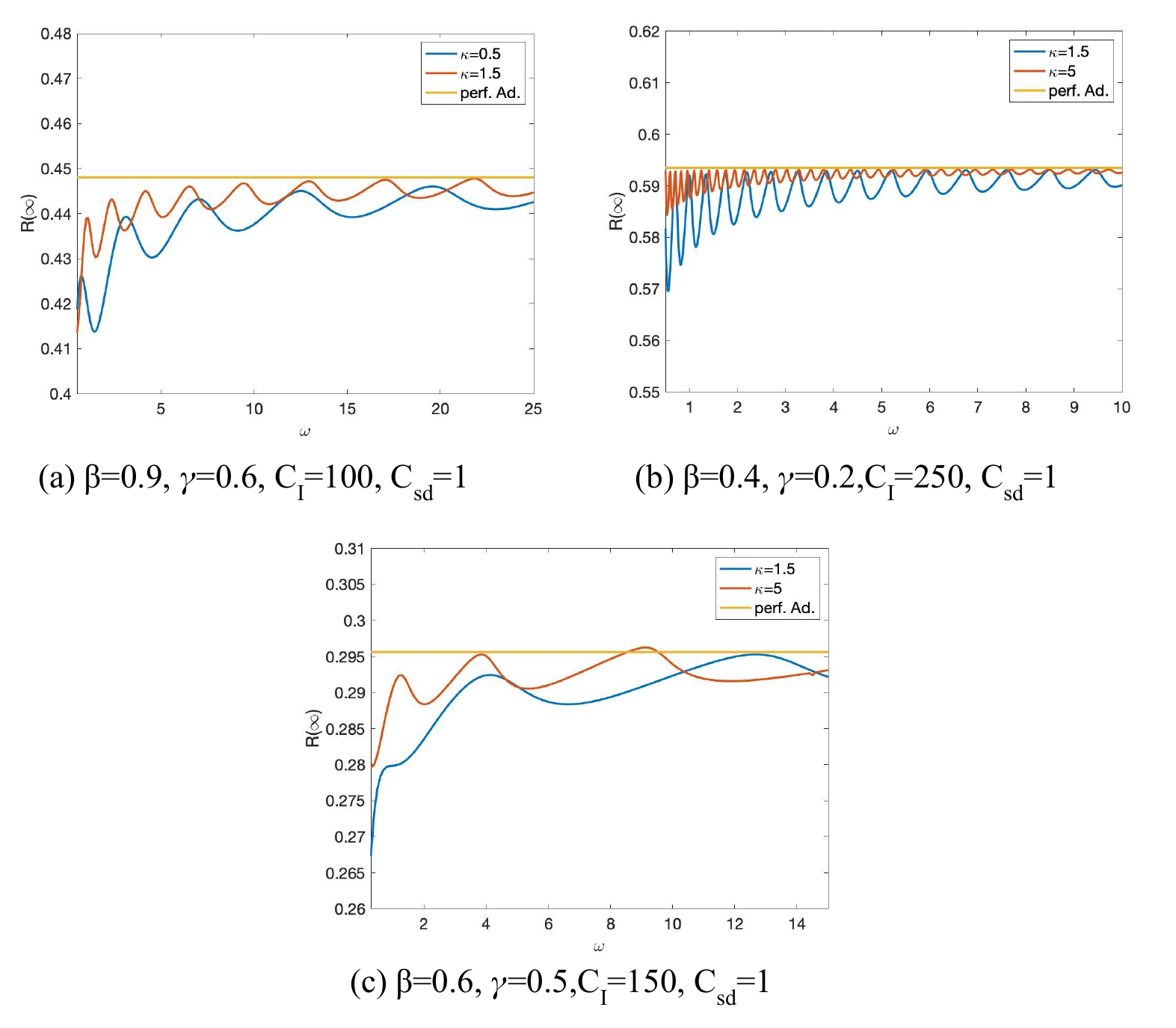}
    \caption{Total number of infections dependent of the responsiveness rate. We compare Model \eqref{eq:SIR_SD} for different rationality parameters with perfect adaption. Again, we have the initial condition $ I_0 = 0.01, \mathcal{E}_0 = 0.2$. $R(\infty)$ shows oscillations with decreasing amplitude converging to perfect adaption here again for the same reasons as explained in Subsection \ref{costs}. This also explains the increasing number of oscillations for increasing $\kappa$ and $C_{\text{I}}$. Moreover, $R(\infty)$ appears to be increasing for increasing $\omega$, in particular, if $\kappa$ and $C_{\text{I}}$ are small.}
    \label{fig:omega}
\end{figure}

 With increasing $\omega$ the deviations of $I$ from $I^*$ are decreasing due to faster adaption. This leads to a decrease in the amplitude of the oscillations in $R(\infty)$ and to $R(\infty)$ leveling off approximately at $R_{\text{PA}}(\infty)$. As explained before, we mostly have $I_{\text{HI}} < I^*$. Therefore, larger deviations from perfect adaption where $I_{\text{HI}} = I^*$ cause a decrease in $R(\infty)$. An example of this behavior is illustrated in Figure \ref{fig:omega}.

\begin{figure}[h!]
    \centering
    \includegraphics[width=4in]{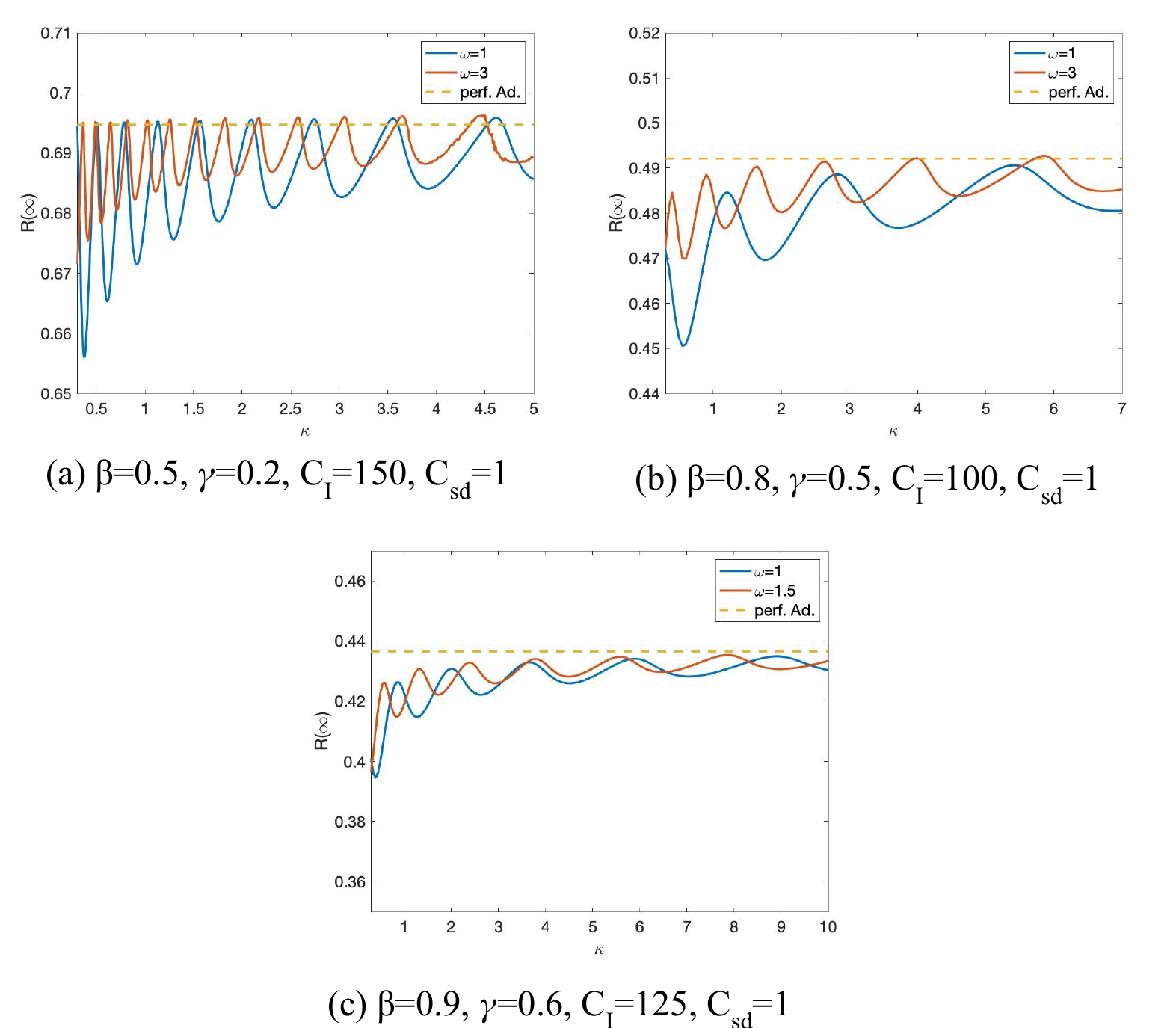}
    \caption{Total number of infections dependent of the rationality parameter. We compare Model \eqref{eq:SIR_SD} with initial condition $I_0 = 0.01, \mathcal{E}_0 = 0.2$ for different responsiveness rates with perfect adaption. Again, we observe the same kind oscillations in $R(\infty)$ that are caused by the development of new waves of infection. Smaller responsiveness parameters $\omega$ cause smaller oscillations and tendentially fewer infections.}
    \label{fig:kappa}
\end{figure}

The rationality parameter $\kappa$ has a nearly similar effect as $\omega$ on the dynamics of our model. In Model \eqref{eq:SIR_SD}, a large rationality parameter $\kappa$ means that individuals change their strategy as soon as the payoff of infection becomes larger than the payoff of social distancing and vice versa.
Therefore, large $\kappa$ induce faster adaption of $\mathcal{E}$ and therefore smaller oscillations of $I$ around $I^*$. An increase in $\kappa$ thus causes a decrease in the duration in the waves of infection as well as smaller maxima and larger minima. Hence, a change in $\kappa$ has a similar effect on $R(\infty)$ as a change in $\omega$. Here as well, we have oscillations caused by the development of new waves of infection, that are decreasing and leveling off at $R_{\text{PA}}(\infty)$. Thus, $R(\infty)$ tends to decrease for smaller $\kappa$ (see Figure \ref{fig:kappa}).

\section{Discussion \& Conclusion}

Social distancing is often used in combination with other control measures such as mask wearing, and testing and isolation. It is worthy of further investigation to account for individual preferences in their adoption choices when multiple interventions for disease mitigation are available. Generally speaking, individuals become less vigilant and feel less need to follow disease intervention measures suggested by public health officials, if the epidemic curve is being bended down, but as a result, the uptick of cases in turn causes individuals to increase their compliance levels. The feedback loop of this sort gives rise to an oscillatory dynamics of behavioral compliance and disease prevalence, as reported in the present work. Similar phenomena have previously been studied in the context of eco-evolutionary dynamics where the payoff structure of individual interactions can be regulated by the environmental feedback~\cite{weitz2016oscillating,tilman2020evolutionary,wang2020steering}. 

Social distancing can be regarded as an altruistic behavior that incurs a cost to oneself but collectively benefits other community members especially these vulnerable in the population. Thus monetary or non-monetary means can be used to incentivize non-compulsory social distancing. For example, during the COVID-19 health crisis, governments have subsidized the cost of staying at home through tax reduction or other stimulus packages for both workers and their employers~\cite{nicola2020socio}.  Besides, an individual who opts for social distancing can create a positive psychological reward, which in fact reduces the perceived overall cost of social distancing. As shown in several experimental works~\cite{capraro2020community,jordan2020community,vanbavel2020community}, encouraging altruistic social distancing, especially if people can afford to do so, through promoting a strong sense of community, empathy and compassion~\cite{van2020using}, can lead to desired compliance of social distancing. In this sense, promoting human cooperation in the social dilemma of disease control is a new promising direction for future work.

While our proof-of-principle model offers enlightening insights into understanding compliance issues in the dilemma of social distancing, \emph{targeted} social distancing can be investigated by further accounting for individual heterogeneity as the attack rate and mortality rate of infectious diseases, such as the influenza~\cite{apolloni2013age,eames2012measured} and the COVID-19 pandemic~\cite{wu2020estimating}, are age-dependent. Thus, extending our model with an age structure will be useful to quantify the heterogeneity in both the risk of infections and the cost of social distancing for each age group. This consideration parameterized using realistic contact mixing matrices in a social network~\cite{block2020social} as well as with an age structure (more generally, multilayer networks~\cite{aleta2020data}) can be used to optimize social network-based distancing protocol (targeted social distancing)~\cite{aleta2020modelling}. Further work along this direction is promising and will help provide practical guidance. Moreover, it appears that instead of the actual likelihood to get infected, one's perceived likelihood to get infected influences the decision whether to engage in social distancing and face covering~\cite{capraro2020community}. Variation in individual risk assessment might therefore influence the results in our model and be an interesting extension to the model in future work.

In sum, we analyze and characterize oscillatory dynamics in the dilemma of social distancing, which arises from the nontrivial feedback between disease prevalence and behavioral intervention. Our results suggest an \emph{oscillatory tragedy of the commons} in disease control when individuals act in their own right without coordination or in the absence of centralized institutions to enforce their compliance, a phenomenon that has been observed in past pandemics like the Spanish flu~\cite{taubenberger2006influenza} and seems to repeat in the current COVID-19 pandemic~\cite{alwan2020covid}. Our work provides new insight into the dual role of human behavior that can fuel, or fight against, the pandemic~\cite{NatHBed}. To resolve the dilemma of disease control from global pandemics to resurgence of common diseases (like measles which has become endemic in some regions~\cite{muscat2009measles}), a deep understanding of pertinent behavioral aspect in disease control and prevention, and large-scale human cooperation in particular, is urgently needed and will help to better inform pandemic support in the future~\cite{van2020using}.

\enlargethispage{20pt}

\section*{Author's contribution}
A.G. \& F.F. conceived the model; A.G. performed simulations and analyses, plotted the figures and wrote the manuscript; F.F. contributed to analyses and writing. All authors give final approval of publication.

\section*{Competing interests}
We have no competing interests to declare.

\section*{Funding}
F.F. is grateful for financial support  by the Bill \& Melinda Gates Foundation (award no. OPP1217336), the NIH COBRE Program (grant no. 1P20GM130454), the Neukom CompX Faculty Grant, the Dartmouth Faculty Startup Fund, and the Walter \& Constance Burke Research Initiation Award.

\section*{Acknowledgements}
We thank Dan Rockmore and Nicholas Christakis for helpful discussions.

\section*{Disclaimer}
We have no disclaimer to disclose.


\end{document}